\documentclass[aps,onecolumn,pra,amsmath,amssymb,floatfix,superscriptaddress]{revtex4}

\usepackage{epsfig}

\usepackage{dcolumn}

\usepackage{bm}

\usepackage{graphicx}

\usepackage{amsmath}

\usepackage{latexsym}

\usepackage{amsfonts}

\usepackage{amssymb}

\usepackage{array}

\usepackage{natbib}

\usepackage{subfig}

\newcommand{\hf}{\frac{1}{2}}

\newcommand{\ket}[1]{|#1\rangle}

\newcommand{\bra}[1]{\langle#1|}

\newcommand{\op}[1]{\hat{#1}}
\newcommand{\opdag}[1]{\hat{#1}^{\dag}}

\begin{document}

\title{Dynamics in a coupled-cavity array}

\author{C. D. Ogden}

\author{E. K. Irish}


\author{M. S. Kim}
\affiliation{School of Mathematics and Physics, Queen's University Belfast, Belfast BT7 1NN, United Kingdom}

\date{\today}



\begin{abstract}
The dynamics of a system composed of two coupled optical cavities, each containing a single two-level atom, is studied over a wide range of detuning and coupling values. A description of the field in terms of delocalized modes reveals that the detuning between the atoms and these modes is controlled by the coupling between the cavities; this detuning in turn governs the nature of the dynamics. If the atoms are highly detuned from both delocalized field modes, the dynamics becomes dispersive and an excitation may be transferred from the first atom to the second without populating the field. In the case of resonance between the atoms and one of the delocalized modes, state transfer between the atoms requires intermediate excitation of the field. Thus the interaction between the two atoms can be controlled by adjusting the coupling between the cavities.
%
\end{abstract}

\maketitle

\section{Introduction}

Cavity quantum electrodynamics (QED) provides a natural setting for distributed quantum information processing (QIP)~\cite{Cirac:1999}. Atoms in highly detuned cavities may be used to store qubits in decoherence-free environments, while photons are well suited to distribute information throughout the system. The strong coupling that can be achieved between atoms and cavity field modes can be used to perform logic operations~\cite{Zubairy:2003, Paternostro:2003}. The physics of cavity QED can also be realized in a solid-state analog known as ``circuit QED,'' in which superconducting qubits are coupled to stripline resonators~\cite{Blais:2004,Wallraff:2004,Schuster:2007}. Circuit QED offers several advantages over traditional cavity QED, including extremely high dipole moments that result in large couplings, a fixed number of qubits in a cavity, and improved scalability. 

One requirement of distributed QIP is the coupling of distant qubits in order to perform state transfer, entanglement generation, or quantum gate operations between separate nodes of the system. In cavity QED, two atoms may be coupled by connecting the cavities in which they are confined via an optical fiber. In an early proposal~\cite{Cirac:1997} for state transfer in this system, the atomic state is converted into a superposition of Fock states of the cavity field mode.  This photonic state then propagates along the fiber to the second cavity, where it is transferred to the second atom. Such a technique, however, is susceptible to decoherence due to losses from the cavity or fiber modes. To decrease these losses, systems of adiabatic transfer have been devised, using a sequence of laser pulses to transfer an atomic state without populating intermediate states. It is possible to minimize the population of the cavities~\cite{Pellizzari:1997} or the fiber~\cite{vanEnk:1999}. Alternatively, the fiber mode can be eliminated from the dynamics if the cavity-fiber coupling is sufficiently strong~\cite{Serafini:2006}. High-fidelity swap and entangling gates may then be produced by turning the atom-cavity interaction on and off. All these proposals, however, still involve excitation of the field in either the cavities or the fiber. 

Field excitation can be avoided entirely by coupling two qubits to a single field mode. In cavity QED, this requires two atoms to be confined within the same cavity~\cite{Zheng:2000}; in circuit QED, two superconducting qubits are coupled to one stripline resonator~\cite{Majer:2007}. If both qubits are highly detuned from the field mode, the interaction with the field mode becomes dispersive. Thus the field mode is effectively eliminated from the qubit dynamics and an excitation may pass between the two qubits without ever populating the field mode. This interaction can produce decoherence-free state transfer and entanglement generation. However, some practical difficulties remain with this scheme. First, the two qubits, interacting with the same field mode, may not be sufficiently independent to act as separate nodes in distributed QIP. Second, there is no way to regulate the qubit-qubit coupling without manipulating the qubits themselves.

Recently, considerable theoretical effort has been devoted to a class of coupled-cavity models that promise to overcome the problem of individual adressability. Inspired by experimental advances in photonic crystals~\cite{Armani:2003}, optical microcavities~\cite{Bayindir:2000}, and superconducting devices~\cite{Wallraff:2004}, these models typically describe a series of optical cavities, each containing one or more atoms or qubits. Photons are permitted to hop between the cavities. Much of the work on coupled-cavity models has focused on similarities to the Bose-Hubbard model~\cite{Fisher:1989} and the prospect of observing quantum phase transitions between Mott insulator and superfluid states~\cite{Hartmann:2006,Greentree:2006,Angelakis:2007b,Huo:2008,Rossini:2007,Hartmann:2008,Hartmann:2007c,Irish:2008b}.

Coupled-cavity models have potential applications in QIP as well, since it should be possible to control and measure individual lattice sites. A proposal for using a coupled-cavity system to generate high-dimensional entangled states of the field was put forward in~\cite{Angelakis:2007a}. Angelakis {\it et al.}~\cite{Angelakis:2008} and Hartmann {\it et al.}~\cite{Hartmann:2007a} have discussed the creation of cluster states for one-way quantum computation. State transfer along an array of polaritonic qubits in a coupled-cavity system was studied in~\cite{Angelakis:2007d}. A protocol for producing two-qubit quantum phase gates in a system of quantum dots coupled to photonic crystal nanocavities was outlined in~\cite{Xiao:2007}. 

Some work has also been done on the specific problem of transferring atomic states between two sites in a coupled-cavity system. {\v S}karja {\it et al.}~\cite{Skarja:1999} studied a system of two weakly coupled cavities. When atoms are passed successively through each cavity, quantum interference effects appear in the final state of the second atom due to the existence of two field modes. The authors focused on the case of exact resonance between the atoms and cavity fields. More recently, Nohama and Roversi~\cite{Nohama:2007} addressed the problem of atomic state transfer in a system of two cavities, each containing a trapped two-level atom. They showed that perfect state transfer occurs periodically, but the analysis was restricted to the case $\Delta = -A$, where $\Delta$ is the detuning between the atom and cavity field and $A$ is the photon hopping strength.  

In the present paper, we examine the dynamics of a simple two-site coupled-cavity model over a large range of values of the qubit-cavity detuning and the photon hopping strength, focusing particularly on atomic state transfer. Although the exact dynamics of the system is rather complicated in the general case, there are several limiting regimes in which it may be simplified. In each of these limits we derive an effective Hamiltonian that captures the essential physical processes and study the resulting dynamics. The limits of large hopping and large detuning each lead to dispersive interactions, allowing state transfer between the atoms without excitation of the field. When the hopping and detuning are approximately equal, a resonant interaction between the atoms and a delocalized mode of the field occurs. Fast state transfer involving excitation of the field is possible in this case. We find, then, that the hopping parameter provides a flexible source of control over the dynamics of the system.

\section{Coupled cavity system}\label{sec:system}

The system under consideration consists of two coupled cavities, each modeled as a single harmonic oscillator mode of frequency $\omega_f$; each cavity contains a single two-level atom with transition frequency $\omega_a = \omega_f + \Delta$~\footnote{Although the system under investigation could be realized in either cavity QED or circuit QED, for the sake of simplicity we will use the language and formalism of cavity QED.}. The atom-cavity interaction is described by the Jaynes-Cummings model~\cite{Jaynes:1963}, using the rotating-wave approximation~\cite{Shore:1993}. The two cavities are coupled so that photons may hop between them. The Hamiltonian governing the system is ($\hbar = 1$)
\begin{equation}
\label{full ham}
\hat{H} = 
\sum^2_{j=1}
   \left[
      \omega_a  \left|e\right>_j\left<e\right|
      +
      \omega_f \hat{a}^{\dagger}_j\hat{a}_j
      +
      g \left(
         \hat{a}_j \hat{\sigma}_{j+} + \hat{a}^{\dagger}_j \hat{\sigma}_{j-}
      \right)
   \right]
   +
   A\left(
      \hat{a}^{\dagger}_1\hat{a}_2 +\hat{a}^{\dagger}_2\hat{a}_1
   \right),
\end{equation}
where $g$ is the atom-cavity coupling strength and $A$ is the cavity-cavity hopping strength. The operator $\op{a}_j$ ($\opdag{a}_j$) is the lowering (raising) operator for the field in cavity $j$. The states $\ket{g}_j$ and $\ket{e}_j$ represent the ground and excited states, respectively, of the atom in cavity $j$. The operator $\hat{\sigma}_{j-}$ ($\hat{\sigma}_{j+}$) is the atomic lowering (raising) operator for cavity $j$. 

The first two terms under the summation in Eq.~\eqref{full ham} correspond to the energy of the bare atoms and field modes, respectively.  The third term represents the transfer of excitations between the atom and field in each cavity. The exchange of photons between cavities is expressed in the final term of Eq.~\eqref{full ham}.

It will prove useful to rewrite the Hamiltonian in terms of delocalized field and atomic modes, which are given by the antisymmetric and symmetric superpositions of the states at the localized sites. The raising operators of the new field modes are given by
\begin{align}\label{deloc op}
\hat{m}_1^{\dagger} &\equiv \frac{1}{\sqrt{2}}
   \left(
      \hat{a}_1^{\dagger} - \hat{a}_2^{\dagger}
   \right),
\nonumber \\
\hat{m}_2^{\dagger} &\equiv 
         \frac{1}{\sqrt{2}}
   \left(
      \hat{a}_1^{\dagger} + \hat{a}_2^{\dagger}
   \right) ;
\end{align}
similarly, the raising operators of the new atomic modes are given by
\begin{align}\label{atomic deloc op}
\hat{s}_{1+} &\equiv \frac{1}{\sqrt{2}}
   \left(
      \hat{\sigma}_{1+} - \hat{\sigma}_{2+}
   \right) ,
\nonumber \\
\hat{s}_{2+} &\equiv \frac{1}{\sqrt{2}}
 \left(
      \hat{\sigma}_{1+} + \hat{\sigma}_{2+}
   \right) .
\end{align}
In terms of these operators, Eq.~\eqref{full ham} becomes
\begin{equation}
\label{full ham m}
\hat{H} = 
\sum^2_{j=1}
   \left[
      \omega_a  \hat{s}_{j+} \hat{s}_{j-} 
      +
      \omega_j \hat{m}_j^{\dagger}\hat{m}_j
      +
      g \left(
         \hat{m}_j \hat{s}_{j+} + \hat{m}^{\dagger}_j \hat{s}_{j-}
      \right)
   \right] .
\end{equation}
In this form, the delocalized atomic mode $s_j$ and the delocalized field mode $m_j$ are governed by a Jaynes-Cummings-type interaction. No coupling occurs between the odd and even modes, so that the system Hamiltonian takes the form of a sum over two independent Jaynes-Cummings Hamiltonians. The transition frequency of the delocalized atomic modes remains unchanged at $\omega_a$. However, the frequencies of the delocalized field modes are shifted from the bare cavity frequency: the frequency of mode $m_j$ becomes $\omega_j$, where $\omega_1 = \omega_f - A$ and $\omega_2 = \omega_f + A$.  The detuning between the atomic mode $s_{j}$ and the field mode $m_j$ is thus given by $\Delta_j$, where 
\begin{align}\label{detunings}
\Delta_1 &= \Delta + A ,
\nonumber \\
\Delta_2 &= \Delta - A .
\end{align}
Our study of the dynamics of the system is focused on the problem of state transfer between the atoms. Atom 1 is prepared in an arbitrary superposition of the ground and excited states, while atom 2 is in the ground state and both field modes are in the vacuum state: 
\begin{equation}\label{initial state}
\ket{\Psi(0)} = 
\left(
   \cos \theta \ket{g}_{1}
   +
   e^{i \phi} \sin \theta \ket{e}_{1}
\right)
\ket{g}_{2} \ket{00} .
\end{equation}
State transfer between the two atoms occurs when atom 2 is in the same superposition of ground and excited states that atom 1 had initially. That is, at some time $t$, the state of the system is given by
\begin{equation}\label{final state}
\ket{\Psi(t)} = 
\ket{g}_{1}\left(
   \cos \theta \ket{g}_{2}
   +
   e^{i \phi} \sin \theta \ket{e}_{2}
\right)
 \ket{00} .
\end{equation}
Since the Hamiltonian conserves total excitation number, the dynamics is restricted to the subspace containing zero or one excitations.  We can thus write the state of the system at time $t$ as $\ket{\Psi (t)} = \ket{\psi_0 (t)} + \ket{\psi_1 (t)}$.  The zero-excitation component $\ket{\psi_0 (t)} = \ket{g}_1 \ket{g}_2 \ket{00}$ is invariant under the action of the Hamiltonian, so we need only consider the dynamics of the single-excitation component $\ket{\psi_1(t)}$.

A general state (not necessarily normalized to $1$) in the single-excitation subspace may be written as
\begin{equation}\label{1 x state}
\ket{\psi_1(t)} = 
a(t)
\ket{g}_1 \ket{g}_2 \ket{10} 
+
b(t)
\ket{g}_1 \ket{g}_2 \ket{01} 
+
c(t)
\ket{e}_1 \ket{g}_2 \ket{00} 
+
d(t)
\ket{g}_1 \ket{e}_2 \ket{00}.
\end{equation}
The coefficients of $\ket{\psi_1(0)}$ that correspond to the initial state in Eq.~\eqref{initial state} are $a(0) = b(0) = d(0) = 0$ and $c(0) = e^{i \phi}\sin \theta$. The conditions for perfect state transfer at time $t$ are, therefore, $a(t) = b(t) = c(t) = 0$ and $d(t) = e^{i \phi}\sin \theta$. 

The time evolution of the general one-excitation state given by Eq.~\eqref{1 x state} may be calculated exactly from the Schr{\"o}dinger equation.  In terms of the delocalized modes $\alpha_j(t) \equiv a(t) + (-1)^j b(t)$ and $\beta_j(t) \equiv c(t) +(-1)^j d(t)$ where $j = {1,2}$, the solutions are given by
\begin{align}\label{full solution}
\alpha_j(t) &=
e^{-i (\omega_a - \Delta_j/2) t}
\left[
   \alpha_j(0) \cos\left( \nu_{j} t \right)
   +
   \frac{i}{\nu_{j}}
   \left(
      \frac{\Delta_j}{2} \alpha_j(0)
      -
      g \beta_j(0)
   \right)
   \sin\left( \nu_{j} t \right)
\right]
\nonumber
\\
\beta_j(t) &=
e^{-i (\omega_a
 -  \Delta_j/2) t}
\left[
   \beta_j(0) \cos\left( \nu_{j} t \right)
   -
   \frac{i}{\nu_{j}}
   \left(
      \frac{\Delta_j}{2} \beta_j(0)
      +
      g \alpha_j(0)
   \right)
   \sin\left( \nu_{j} t \right)
\right],
\end{align}
where 
\begin{equation}
\nu_{j} = 
\sqrt{
	\left(
		\frac{\Delta_j}{2}
	\right)^2 
	+ g^2
}.
\end{equation}
These equations describe complicated dynamics. However, there are several limiting regimes in which the Hamiltonian and the resulting dynamics can be substantially simplified. This is most readily seen in the delocalized basis. It is evident from Eq.~\eqref{detunings} that if, for example, $\Delta = A$, $\Delta_2 = 0$ and the atomic mode $s_2$ is on resonance with the field mode $m_2$. However, provided that $\{\Delta, A\} \gtrsim g$, the modes $s_1$ and $m_1$ will be off-resonant and will not contribute significantly to the dynamics on short time scales. Away from the resonances at $\Delta = \pm A$, both atomic modes are detuned from their corresponding field modes and the system dynamics becomes dispersive in nature.

We consider three limiting cases: large hopping ($A \gg \{\Delta,g\}$), large detuning ($\Delta \gg \{A,g\}$), and near-resonance ($\Delta + A \gg \{\Delta - A, g\}$). For each, we determine the effective dispersive-case Hamiltonian, which illustrates the important physical processes, and then calculate the explicit single-excitation dynamics from the full solution given in Eq.~\eqref{full solution}.

%
%

\section{Large hopping}\label{sec:large hopping}

The first limit we examine is the large-hopping limit, in which $A \gg \{\Delta , g\}$. The atoms are highly detuned from both delocalized field modes ($\{\Delta_1,\Delta_2 \} \gg g$). Transitions in which an excitation is passed from atom to field (or vice versa) involve the exchange of a large amount of energy and so are strongly supressed. We can simplify the dynamics by constructing an effective Hamiltonian in which such transitions are absent. 

We begin by switching to the interaction picture with respect to 
\begin{equation}
\hat{H}_0 = 
\sum^2_{j=1}
   \left(
      \omega_a  \left|e\right>_j\left<e\right|
      +
      \omega_f \hat{a}^{\dagger}_j\hat{a}_j
   \right)
   +
   A\left(
      \hat{a}^{\dagger}_1\hat{a}_2 +\hat{a}^{\dagger}_2\hat{a}_1
   \right) .
\label{H 0}
\end{equation}
In this picture, the Hamiltonian becomes
\begin{equation}
\label{cc int ham}
\hat{H}_I = g
\left[
   \hat{m}_1^{\dagger}\hat{s}_{1-} e^{-i\Delta_1 t}
   +
   \hat{m}_2^{\dagger}\hat{s}_{2-} e^{-i\Delta_2 t}
   + h.c.
\right].
\end{equation}
Since $\Delta_1$ and $\Delta_2$ are both taken to be large in this limit, the terms in this interaction Hamiltonian rotate rapidly in time. As in the standard rotating-wave approximation~\cite{Jaynes:1963,Shore:1993}, the time average of such rapidly rotating terms may be set to zero, thus eliminating the highly off-resonant transitions between atomic and field modes.

To simplify the Hamiltonian in this manner, without entirely eliminating the atom-field interaction terms, we follow the method of James and Jerke~\cite{James:2007}. The authors define an effective Hamiltonian $\hat{\mathcal{H}}_{eff}$ from the time derivative of the time-averaged evolution operator,
\begin{equation}\label{non-hermitian eff ham}
i\frac{\partial}{\partial t}
\overline{
   \hat{U}(t)
}
=
\hat{\mathcal{H}}_{eff}
\overline{
   \hat{U}(t)
} ,
\end{equation}
where the time average of the operator $\hat{\mathcal{O}}(t)$ is defined as 
\begin{equation}
\overline{
   \hat{\mathcal{O}}(t)
} = 
\int^{\infty}_{-\infty}
   f\left( t - t' \right)
   \hat{\mathcal{O}}(t')
dt'.
\end{equation}
Here, the probability density function $f(t-t')$ acts to remove the high frequency components of the operator.

By comparing Eq.~\eqref{non-hermitian eff ham} with the Schr{\"o}dinger equation for the non-time-averaged evolution operator, 
\begin{equation}
i\frac{\partial}{\partial t}
\hat{U}(t)
=
\hat{H}_I \hat{U}(t) ,
\end{equation}
we can write the effective Hamiltonian as a series in $\hat{H}_I$ and $\hat{U}$.  For a system with a weak interaction, we discard terms beyond second order in $\hat{H}_I$. The Hermitian part~\footnote{The effective Hamiltonian defined by Eq.~\eqref{non-hermitian eff ham} is non-Hermitian and thus gives rise to non-unitary dynamics. It can be shown~\cite{James:2007} that the unitary dynamics is given by the Hermitian part of the effective Hamiltonian.} of the effective Hamiltonian may thus be written as 
\begin{equation}
\label{eff ham u}
\hat{H}_{eff} = \overline{\hat{H}_I(t)}
+\hf\left(
   \overline
   {
      \left[
         \hat{H}_I(t),\hat{U}_1 (t)
      \right]
   }
   -
   \left[
      \overline
      {
         \hat{H}_I(t)
      },
      \overline
      {
         \hat{U}_1 (t)
      }
   \right]
\right) .
\end{equation}
The term $\hat{U}_1 (t)$, given by 
\begin{equation}
\label{evo op}
\hat{U}_1 (t) = 
\frac{1}{i} 
\int^t_{t_0}
   \hat{H}_I (t')
dt',
\end{equation}
is the first order term in the expansion in $\hat{H}_I(t)$ of the evolution operator $\hat{U}(t)$.

Applying this method to our system, we replace $\hat{U}_1(t)$ in Eq.~\eqref{eff ham u} by
\begin{equation}
\label{v}
\hat{V}(t) = g
\left[
   \frac{\hat{m}_1^{\dagger}\hat{s}_{1-} e^{-i\Delta_1t}}
   {\Delta_1}
   +
   \frac{\hat{m}_2^{\dagger}\hat{s}_{2-} e^{-i\Delta_2t}}
   {\Delta_2}
   - h.c.
\right] ,
\end{equation}
where $i\hat{V}(t)$ is the time integral of the interaction Hamiltonian.

The effective Hamiltonian can thus be determined by taking the time-average of \eqref{cc int ham}, \eqref{v}, and 
\begin{equation}
\begin{split}\label{hiv}
\left[
   \hat{H}_I(t),\hat{V} (t)
\right]
&=
\frac{2g^2}{\Delta_1}
\left[
   \hat{m}_1\hat{s}_{1+}
   ,
   \hat{m}_1^{\dagger}\hat{s}_{1-}
\right]
+
\frac{2g^2}{\Delta_2}
\left[
   \hat{m}_2\hat{s}_{2+}
   ,
   \hat{m}_2^{\dagger}\hat{s}_{2-}
\right]
 \\
&
+ 
g^2 e^{2iAt}
\left(
   \frac{1}{\Delta_1}
   +
   \frac{1}{\Delta_2}
\right)
\left[
   \hat{m}_1\hat{s}_{1+}
   ,
   \hat{m}_2^{\dagger}\hat{s}_{2-}
\right]
+
g^2 e^{-2iAt}
\left(
   \frac{1}{\Delta_1}
   +
   \frac{1}{\Delta_2}
\right)
\left[
   \hat{m}_2\hat{s}_{2+}
   ,
   \hat{m}_1^{\dagger}\hat{s}_{1-}
\right] .
\end{split}
\end{equation}
In the present limit, those terms rotating at rate $\Delta_1$, $\Delta_2$, or $2A$ are considered to be fast, so their time averages may be set to zero. The effective Hamiltonian is thus given by the first two terms of Eq.~\eqref{hiv}, divided by 2. Returning to the Schr{\"o}dinger picture, the effective Hamiltonian becomes
\begin{equation}
\hat{H}'
=
\hat{H}_0 
+
\sum^2_{i,j=1}
   \frac{g^2}{2 \Delta}_i
   \left(
      \hat{m}_i \hat{m}_i^{\dagger}
      \ket{e}_j\bra{e}
      -
      \hat{m}_i^{\dagger}\hat{m}_i
      \ket{g}_j\bra{g}
   \right)
+
G
A
\left(
   \hat{\sigma}_{1+}\hat{\sigma}_{2-}
      +
      \hat{\sigma}_{2+}\hat{\sigma}_{1-}
\right) ,
\end{equation}
where 
\begin{equation}\label{capital G}
   G \equiv \frac{g^2}{\Delta_1 \Delta_2}.
\end{equation}
This Hamiltonian has three contributing terms.  The first, $\hat{H}_0$, is the energy of the bare atomic and delocalized field modes, given by Eq.~\eqref{H 0}.  The summation represents a dispersive interaction between the atoms and the delocalized cavity modes $m_1$ and $m_2$. As with dispersive interactions between atoms and single cavities, there is no transfer of energy; the atoms, however, experience a Stark shift, the magnitude of which depends on the population of the field modes. Finally, the last term describes a transfer of energy between the two atoms without excitation of the field modes.

The behavior of two atoms coupled to two highly detuned, non-interacting field modes described here closely resembles the behaviour of two atoms coupled to a single detuned field mode~\cite{Zheng:2000}. In the single-mode case, the atoms are Stark-shifted by the field mode, with the shift proportional to $g^2 / \Delta$ (where $\Delta$ is the detuning between the atom and the field mode). In the two-mode case, the atoms experience a shift due to each mode, proportional to $g^2 / \Delta_i$. The total effect is the average of the two single-mode shifts. In both cases the interatomic energy transfer occurs without excitation of the cavity modes. The rate of transfer, however, differs between the two models. In the one-mode model, the rate is inversely proportional to the atom-field detuning. In the two-mode model, it is inversely proportional to the product $\Delta_1\Delta_2$ of the detunings between the two field modes and the atoms, and proportional to the intercavity hopping $A$.

To find a more precise description of the system's behavior, we turn to the equations of motion given by Eq.~\eqref{full solution}. Taking the limit of large hopping, we obtain
\begin{align}\label{evol A}
a(t) &=
e^{-i
   \left(
      \omega_f -G\Delta
   \right)
t}
\left[
   a(0)\cos \left(A t\right)
   -
   ib(0) \sin(At)
\right]
\nonumber \\
b(t) &=
e^{-i
   \left(
      \omega_f -G\Delta
   \right)
t}
\left[
   - ia(0) \sin(At)   
   +
   b(0) \cos \left(A t\right)
\right]
\nonumber \\
c(t) &=
e^{-i
   \left(\omega_a +G\Delta\right)
t}
\left[
   c(0) \cos \left( GAt \right)
   -
   id(0) \sin \left( GAt \right)
\right]
\nonumber \\
d(t) &=
e^{-i
   \left(\omega_a +G\Delta\right)
t}
\left[
   -ic(0) \sin \left( GAt \right)
   +
   d(0) \cos \left( GAt \right)
\right].
\end{align}
As expected, there is no transfer of energy from field to atom or vice versa.  If the excitation is initially in one of the field modes, it will simply oscillate between the cavities at rate $A$; this is the same as the rate of intercavity transfer in the absence of atoms. If the excitation is initially in one of the atoms, it will oscillate between the two cavities at rate $GA$, without populating the field modes.  Thus, at times $t= \left( n + \hf \right)\pi/GA$, an excitation initially in atom 1 will be fully transferred to atom 2.  Increasing the atom-cavity detuning $\Delta$ increases this rate of transfer, as doing so decreases the product $\Delta_1 \Delta_2$ of the detuning between the atoms and the two delocalized field modes, and thus increases $G$.  Increasing the hopping strength $A$ causes a quadratic decrease in $G$.  This is sufficient to overcome the linear scaling of the transfer rate $GA$ with $A$; the transfer rate therefore diminishes with increasing hopping strength.

\begin{figure}
\includegraphics[scale=1]{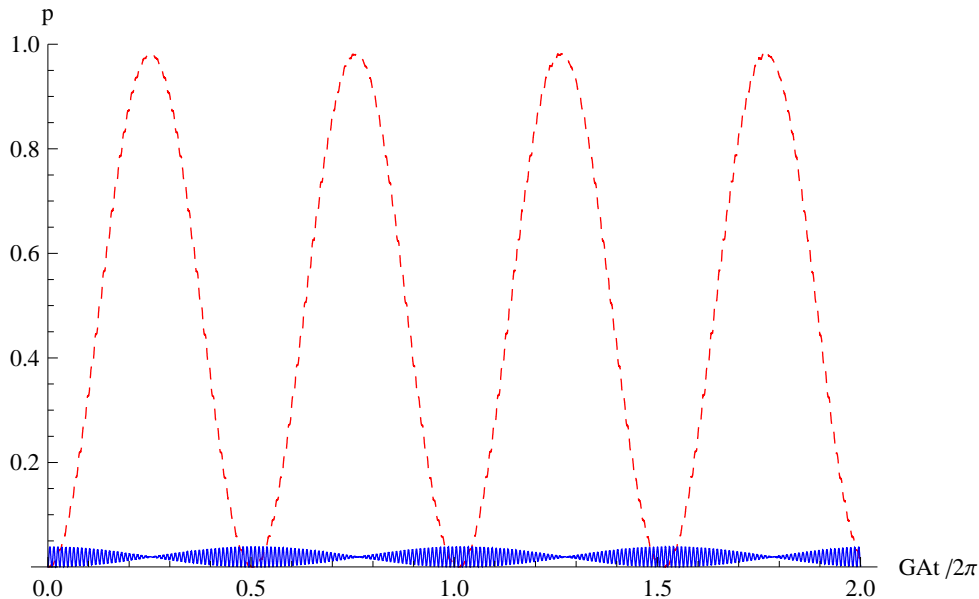}
\caption{\label{fig:hi A}(Color online) Dynamics of atomic and field excitation in the case $A \gg \Delta$, calculated by numerical solution of the full Hamiltonian, Eq.~\eqref{full ham}. The initial state has a single excitation in atom 1. The dashed red curve gives the probability, as a function of time, of finding the excitation in atom 2.  The solid blue curve corresponds to the probability of excitation of the field modes. The system parameters are $A = 10g$, $\Delta = 0.1g$, and $\omega_f = 1000g$.}
\end{figure}

Figure~\ref{fig:hi A} shows the evolution of the system in the large hopping limit, calculated numerically from the full Hamiltonian in Eq.~\eqref{full ham}. The initial state has a single excitation localized in atom 1. In agreement with the approximate solution of Eq.~\eqref{evol A}, the probability of finding the excitation in atom 2 undergoes sinusoidal oscillation at a rate $\simeq GA$, and the probability of finding the excitation in either cavity mode is almost zero.

Having seen how a general state in the single-excitation subspace behaves, we turn to the problem of transferring a qubit state from atom 1 to atom 2. The initial state is that given in Eq.~\eqref{initial state}, which consists of a superposition of zero- and one-excitation states. As the zero-excitation state is unaffected by the Hamiltonian, we need only consider the evolution of the single-excitation part. In terms of the general state of Eq.~\eqref{1 x state}, the initial conditions are $a(0) = b(0) = d(0) = 0$; the condition for successful state transfer at time $\tau$ is $d(\tau) = c(0)$. This places two requirements on the interaction time $\tau$. First, to ensure that the final state of atom 2 has the correct ratio between the excited and ground states, we need $GA\tau = (2n + 1/2)\pi$ ($n=0,1,2,\dots$). Second, $\tau$ must be chosen such that $(\omega_a + G\Delta)\tau = (2m - 1/2)\pi$ ($m=0,1,2,\dots$) in order for the phase difference between the two states to be correct. Both can be achieved if the system parameters are selected so that
\begin{equation}
\frac{\omega_a + G\Delta}{GA} = \frac{4m - 1}{4n+1}.
\end{equation}
If only the first condition can be met, then, provided that the interaction time is accurately known, a single qubit operation can be applied to atom 2 to correct the phase difference and complete the state transfer.

This transfer of states between atoms in separate cavities without populating the cavity modes has significant practical benefits. The method of transferring the atomic state to the field and then from the field to the second atom~\cite{Cirac:1997} is susceptible to decoherence from photon loss in the cavities or optical fibers. Qubit-qubit excitation transfer without populating the field modes, as in the proposal by Zheng and Guo~\cite{Zheng:2000} and the realization by Majer {\it et al}.~\cite{Majer:2007}, minimizes decoherence due to photon loss. Furthermore, it is possible to extend this technique to decoherence-free entanglement generation, as suggested in~\cite{Zheng:2000}. For an initial state with a single excitation in atom 1 [$a(0) = b(0) = d(0) = 0$, $c(0) = 1$], Eq.~\eqref{evol A} shows that the two atoms will be maximally entangled at times $\tau$ such that $\tau = \left( n + \frac{1}{2} \right) \pi/2GA$.

\section{Large Detuning}

We next consider the limit of large detuning, in which $\Delta \gg \{A, g\}$.  As before, the atoms are highly detuned from both delocalized field modes, so atom-field transitions may be eliminated. The analysis proceeds similarly to that in the large hopping limit. This time, however, $A$ is a small quantity and thus only those terms that rotate at rate $\Delta_1$ or $\Delta_2$ are eliminated from the time averages of Eqs. \eqref{cc int ham}, \eqref{v}, and \eqref{hiv}. The resulting Hamiltonian is, in the Schr{\"o}dinger picture,
\begin{equation}\label{mod ham lo A}
\hat{H}'=
\hat{H}_0
+
   \sum^2_{j=1}
   G \Delta
   \left(
      \hat{a}_j\hat{a}_j^{\dagger} \ket{e}_j\bra{e}
      -
      \hat{a}_j^{\dagger}\hat{a}_j \ket{g}_j\bra{g}
   \right)
   +
   \frac{GA}{2}
   \left(
      \hat{m}_2^{\dagger}\hat{m}_2 - \hat{m}_1^{\dagger}\hat{m}_1
   \right)
   \left(   
      \hat{\sigma}_{1z} + \hat{\sigma}_{2z}
   \right)
   +
   G A 
   \left(
      \hat{\sigma}_{1+}\hat{\sigma}_{2-}
      +
      \hat{\sigma}_{2+}\hat{\sigma}_{1-}
   \right) .
\end{equation}
The atom-atom excitation transfer, given by the final term of Eq.~\eqref{mod ham lo A}, is identical to that found in the previous limit. The Stark shift terms, however, take a different form. In the large-hopping limit, the shift for each atom was determined purely by its interaction with the two delocalized field modes. In the current limit, with the atom-cavity detuning much greater than the hopping strength, the shift splits into two contributing factors. The first of these, proportional to $G \Delta$, is due to the local interaction between each atom and the field mode contained within its cavity. (Note that if $A=0$ the Hamiltonian reduces to that of two isolated far-off-resonance atom-cavity systems.) The second contribution, proportional to $GA / 2$, is due to an interaction between the atoms and the delocalized field modes. The Stark shift due to this term (which, as $A \ll \Delta$, is much smaller than the local effect) is proportional to the difference between the populations of the two field modes.


The evolution of a single-excitation state reduces to the same equations of motion as in the limit $A \gg \Delta$, given by Eq.~\eqref{evol A}. As before, there is no transfer of energy between the field modes and the atoms, and an excitation that is initially in an atom will oscillate between the two atoms at rate $GA$. In contrast with the previous limit, increasing the hopping strength decreases the product $\Delta_1 \Delta_2$, thus increasing both $G$ and the interatomic excitation transfer rate. Conversely, increasing $\Delta$ decreases the transfer rate.

\section{Comparable Hopping and Detuning}

Finally, we turn to the limit in which the atoms are nearly resonant with one of the delocalized field modes.  Unlike the previous two limits where the atoms were highly detuned from both modes, direct transfer of energy takes place between the atoms and the field, creating a different mechanism for interatomic energy transfer. Specifically, we consider the limit $\Delta + A \gg \{|\Delta -A|,g\}$, with $\Delta > 0$. In this case the atoms are nearly resonant with the field mode $m_2$ but highly detuned from $m_1$.  

For simplicity, we do not consider the equivalent parameter regime for negative detuning, i.e. $A - \Delta \gg \{|\Delta +A|,g\}$. In this limit, the atoms are nearly resonant with $m_1$ and highly detuned from $m_2$. Unlike our previous work on the ground state of the coupled cavity system, which showed a marked asymmetry between positive and negative detuning~\cite{Irish:2008b}, the dynamics depends only on the absolute value of the detuning and thus the positive- and negative-detuning cases exhibit the same behavior.



We derive the effective Hamiltonian by eliminating terms rotating at $2A$ and $\Delta_1$ from the time averages of Eqs. \eqref{cc int ham}, \eqref{v}, and \eqref{hiv}. The near-resonance condition means that $\Delta_2$ is small and the corresponding terms must be retained. Returning to the Schr{\"o}dinger picture, we find
\begin{equation}\label{mod ham del A}
\begin{split}
\hat{H}' &= \hat{H}_0 + 
g
\left[
   \hat{m}_2^{\dagger}
   \hat{s}_{2-}
   +
   \hat{m}_2
   \hat{s}_{2+}
\right]
\\
&+
   \frac{g^2}{2 \Delta_1}
   \left[
      \hat{m}_1\hat{m}_1^{\dagger}
      \left(
      	\ket{e}_1\bra{e} + \ket{e}_2\bra{e}
      \right)
      -
      \hat{m}_1^{\dagger}\hat{m}_1
      \left(
      	\ket{g}_1\bra{g} + \ket{g}_2\bra{g}
      \right)
      -
      \left(
         \hat{\sigma}_{1+}\hat{\sigma}_{2-}
         +
         \hat{\sigma}_{2+}\hat{\sigma}_{1-}
      \right)
   \right].
\end{split}
\end{equation}
The second term in this Hamiltonian describes the transfer of energy between the near-resonant atomic and field modes $s_{2}$ and $m_2$. The third term represents the dispersive interaction between modes $s_{1}$ and $m_1$. As in the previous limits, this is characterized by a Stark shift of the atoms (which depends on the population of $m_1$) as well as a direct transfer of energy between the atoms without stimulating the field modes~\footnote{Indeed, this term is, but for a factor of two, identical to the case of two atoms interacting with a single detuned field mode~\cite{Zheng:2000}.}. As $g \gg g^2/\Delta_1$, the evolution is dominated by the near-resonant term in Eq.~\eqref{mod ham del A}, so the dispersive term may be neglected for times $t \ll \Delta_1 / g^2$.

To study the dynamics in detail,  we turn to the equations of motion for the delocalized modes given in Eq.~\eqref{full solution}. We apply the current limit, $\Delta + A \gg \{|\Delta -A|,g\}$, and make the additional assumption that modes $s_{2}$ and $m_2$ are very nearly on resonance, i.e. $|\Delta - A| \ll g$.  If the field modes are initially unpopulated, i.e. $\alpha_j(0) = 0$, the evolution of the atomic modes is given by
\begin{align}\label{the betas}
\beta_1(t) &=
   e^{-i 
     (
         \omega_a
         + g^2/\Delta_1
      )
   t}
   \beta_1(0) ,
\nonumber \\
\beta_2(t) &=
   e^{
      -i (
         \omega_a - \Delta_2/2
      )t
   }
   \beta_2(0) \cos 
   \left[
      \left(
         g + \frac{\Delta_2^2}{8g}
      \right)
      t
   \right] .
\end{align}
The population of atomic mode $s_{2}$, represented by coefficient $\beta_2(t)$, oscillates at rate $\simeq g$ due to the excitation transfer between $s_{2}$ and field mode $m_2$.  Mode $s_{1}$, represented by $\beta_1(t)$, has a constant population because it is coupled only dispersively to the field. The phases of the two modes rotate at different rates, with the difference $\delta$ between them given by
\begin{equation}
\delta = \frac{\Delta_2}{2} + \frac{g^2}{\Delta_1}.
\end{equation}
The second term in this expression corresponds to the Stark shift of atomic mode $s_{1}$ due to its dispersive interaction with field mode $m_1$. 

Let us consider the case of exact resonance, $\Delta_2 = 0$. If the effect of the dispersive processes is negligible, i.e. $g^2 / \Delta_1 \ll \pi/t$, so that $\delta \simeq 0$, the two delocalized atomic modes rotate at approximately the same rate. The behavior of the localized atom and cavity modes, in this case, reduces to simple sinusoidal oscillation.  Taking the initial conditions $a(0) = b(0) = d(0) = 0$, we find that
\begin{align}\label{delta A resonant}
a(t) &= b(t) =
   -\frac{i c(0)}{2} 
   e^{-i\omega_a t}
   \sin (gt) ,
\nonumber \\
c(t)&=
   c(0)   
   e^{-i \omega_a t}
   \cos^2 \left( \frac{gt}{2} \right) ,
\nonumber \\
d(t)&=
   c(0)
   e^{-i \omega_a t}
   \sin^2 \left( \frac{gt}{2} \right),
\end{align}
in agreement with the results of Nohama and Roversi~\cite{Nohama:2007}. Excitation transfer between atoms 1 and 2 occurs at times $\tau = (2n + 1)\pi/g$ with $n=0,1,2,\dots$. In contrast with previous limits, an excitation that is initially in an atomic mode will be passed to the field. This direct transfer of energy from atom 1 to the field, then from the field to atom 2, is responsible for the high rate of interatomic transfer. Though the transfer occurs much faster than in the dispersive limit, the speedup comes at the cost of exposing the system to decoherence via cavity loss.

\begin{figure}
\includegraphics[scale=1]{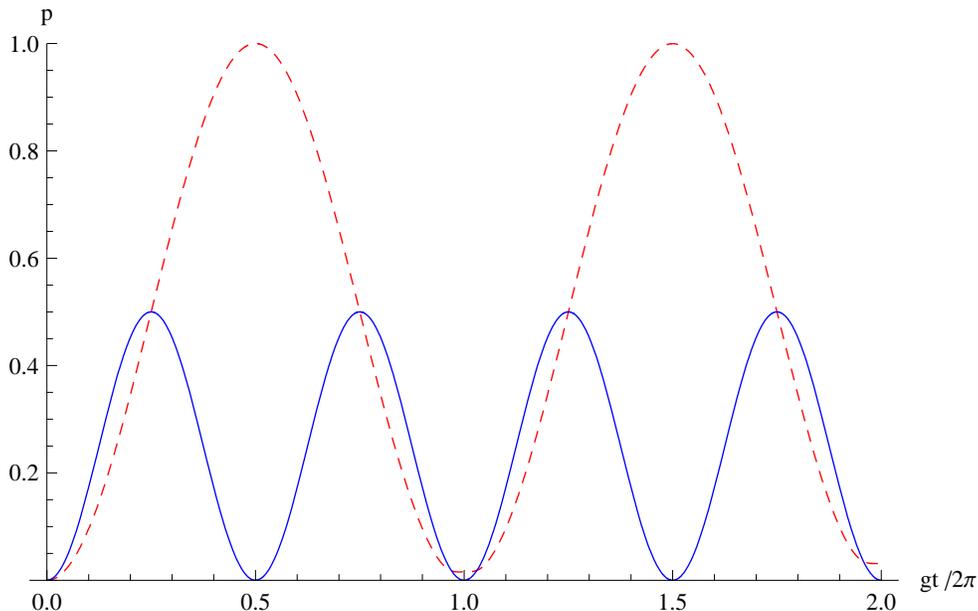}
\caption{\label{fig:hihidethop} Dynamics of atomic and field excitation in the resonant case $A = \Delta$, calculated by numerical solution of the full Hamiltonian, Eq.~\eqref{full ham}. The initial state has a single excitation in atom 1. The dashed red curve gives the probability, as a function of time, of finding the excitation in atom 2.  The solid blue curve corresponds to the probability of excitation of the field modes. The system parameters are $A = 100g$, $\Delta = 100g$, and $\omega_f = 1000g$.}
\end{figure}

Figure~\ref{fig:hihidethop} shows the numerically determined evolution of the system for the case of exact resonance. The probability of finding an excitation in atom 2 oscillates between approximately zero and unity at rate $g$.  We note, however, that this probability does not return to exactly zero.   This is due to the Stark shift of mode $s_{1}$. As discussed below, this shift becomes significant for large interaction times and causes beating in the evolution of the populations of atoms 1 and 2.  Additionally, in contrast with the previous limits, the field modes become populated; the probability of finding a photon in either cavity mode oscillates between zero and 0.5.

Perfect state transfer requires that $d(\tau) = c(0)$, introducing an additional constraint on the interaction time $\tau$. As well as selecting $\tau$ so that $\tau = (2n + 1)\pi/g$, it is also necessary to have $\omega_a \tau = 2m\pi$. As shown in~\cite{Nohama:2007}, these can be simultaneously achieved by setting $\omega_a = 2 l g$ where $l$ is an integer. If this is impractical, the interaction time may be chosen such that $\tau = (2n + 1)\pi/g$, so the final state of atom 2 has the correct population; the phase difference between excited and ground states can then be corrected by means of a single-qubit operation.

\begin{figure}
\includegraphics[scale=1]{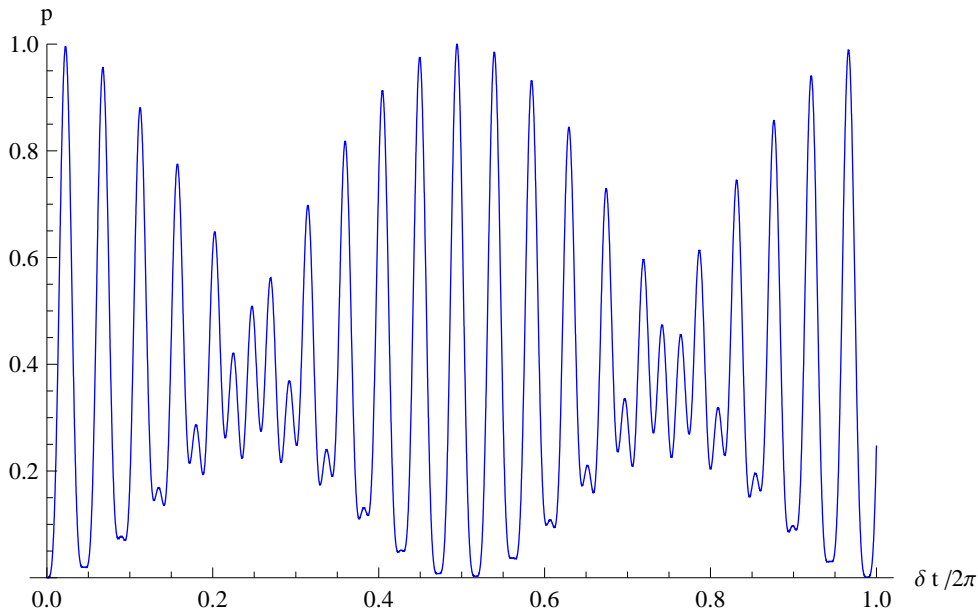}
\caption{\label{fig:hihidethophop}(Color online) Probability of excitation transfer in the near-resonant case, calculated by numerical solution of the full Hamiltonian, Eq.~\eqref{full ham}. The initial state has a single excitation in atom 1. The  curve shows the probability of finding the excitation in atom 2. The system parameters are $A = 100.1g$, $\Delta = 100g$, and $\omega_f = 1000g$.}
\end{figure}

Away from exact resonance, $\delta \ne 0$ and the phases of the two delocalized atomic modes no longer rotate at the same rate. The competition between the resultant oscillating phase difference and the excitation transfer between $s_{2}$ and $m_2$ leads to beating behavior in the evolution of the local atomic modes. Again taking the initial conditions $a(0) = b(0) = d(0) = 0$, we obtain from Eq.~\eqref{full solution} the expressions
\begin{align}
a(t) &= b(t) =
   -\frac{ic(0)}{2}
   e^{-i(\omega + \Delta_1/2)t}
   \sin
   \left[
      \left(
         g + \frac{\Delta_2^2}{8g}
      \right)
      t
   \right] ,
\nonumber \\
c(t)&=
   \frac{
      e^{-i(\omega + \Delta_1/2)t}
   }{2}
   \left\{
      \cos
      \left[
         \left(
            g + \frac{\Delta_2^2}{8g}
         \right)
         t
      \right]
      +
      e^{-i \delta t} 
   \right\} ,
\nonumber \\
d(t)&=
   \frac{
      e^{-i(\omega + \Delta_1/2)t}
   }{2}
   \left\{
      \cos
      \left[
         \left(
            g + \frac{\Delta_2^2}{8g}
         \right)
         t
      \right]
      -
      e^{-i \delta t}
   \right\}
.
\end{align}
Figure~\ref{fig:hihidethophop} shows the probability of transfer when the atoms are detuned from field mode $m_2$ by $0.1g$. The fast oscillation occurs at a rate $\simeq g$, as in the resonant case. For the first Rabi oscillation, the probability of transferring an excitation from atom 1 to atom 2 is nearly unity.  The effect of the detuning is to reduce the transfer probability over subsequent cycles; the maximum probability oscillates at rate $\delta$.  

This beating can be avoided for certain values of the system parameters. The difference $\delta$ between the rates of rotation of the atomic modes $s_{1}$ and $s_{2}$, has two contributing factors: the detuning between $s_{2}$ and $m_2$, and the Stark shift of mode $s_{1}$. If these two factors have equal and opposite effects so that $\delta = 0$, the evolution of the system will reduce to the simple oscillation described by Eq.~\ref{delta A resonant}.  Conversely, beating will occur even in the resonant case on timescales that are sufficiently long for the dispersive processes to become significant, i.e. $t \simeq 1/\delta = \Delta_1/g^2$.

\section{Discussion}

We have studied the dynamics of a system consisting of two cavities, each interacting with a two-level atom with coupling strength $g$ and detuning $\Delta$. The cavities are coupled through a photon hopping term with strength $A$. The electromagnetic field in such a system may be described either in terms of local cavity fields or delocalized modes. The behavior of the system is governed by the relative values of the three parameters $g$, $\Delta$, and $A$. Let us assume that the values of $g$ and $\Delta$ have already been set. Three different regimes of behavior may then be identified, depending on the value of $A$. When $A=0$, the two atom-cavity systems evolve independently. In the limits $A \ll \Delta$ and $A \gg \Delta$, the field may be eliminated from the dynamics and the atoms interact directly. When $A = \pm \Delta$, the atoms interact resonantly with one of the delocalized field modes.

This variety of interaction, governed by the single parameter $A$, could allow coupled cavities to form a useful and highly adaptable element of a QIP system.
By selecting different values for the atom-cavity detuning and the hopping strength, it is possible to choose the properties of the atomic state transfer. In the dispersive limit, the transfer will take place slowly, but since the field modes are never excited the process is protected against decoherence. When the atoms are on resonance with one delocalized field mode, the transfer is more rapid, although it is subject to decoherence due to cavity loss. 

Such dispersive and resonant methods of state transfer could also be realized in a system of two atoms interacting with a single cavity. Coupled cavity systems, however, offer additional advantages. Qubits coupled to separate cavities could be more easily isolated from each other, eliminating cross-talk. If the cavity-cavity coupling is turned off, the two qubit-field systems become independent, and thus able to act as part of separate QIP subsystems. Furthermore, by controlling the hopping strength, the interaction between the qubits can be controlled without manipulating the qubits themselves.  

There are several possibilities for experimental realizations of coupled-cavity systems. Within the cavity QED setting, high Q-factor cavities could potentially be coupled by means of an imperfectly reflecting mirror. Atoms would be either trapped in the cavities or passed simultaneously through them. Another potential setting is that of semiconductor microcavites. A system of two coupled microdisk resonators, each with an embedded quantum dot, has already been demonstrated~\cite{Benyoucef:2008}. Perhaps the most promising system, though, is circuit QED. Experimental work on single qubit-stripline systems is already well established~\cite{Wallraff:2004,Schuster:2007}. Majer {\it et al.} have demonstrated that it is possible to couple two superconducting qubits via a single resonator without significantly populating the resonator modes. Furthermore, a scheme for controllable coupling between stripline resonators has been proposed recently~\cite{Hu:2007}. A pair of qubit-stripline systems, connected by a controllable coupling element, would provide a flexible and realistic experimental implementation of the coupled-cavity model.

\acknowledgments

We would like to thank G.~S. Agarwal for helpful discussions. This work was supported by UK EPSRC and QIPIRC. 


\end{document}